# Twist Engineering of Anisotropic Excitonic and Optical Properties of a Two-Dimensional Magnetic Semiconductor

Qiuyang Li,[1] Xiaohan Wan,[1] Senlei Li,[2] Adam Alfrey,[3] Wenhao Liu,[4] Zixin Zhai,[4] Wyatt Alpers,[1] Yujie Yang,[1] Irmina Wladyszewska,[1] Christiano W. Beach,[1] Liuyan Zhao,[1] Bing Lv,[4] Chunhui Rita Du,[2] Kai Sun,[1] Hui Deng[1,3,*]

[1]Department of Physics, University of Michigan, Ann Arbor, Michigan 48109, United States

[2]School of Physics, Georgia Institute of Technology, Atlanta, GA, 30332, United States

[3]Applied Physics Program, University of Michigan, Ann Arbor, Michigan 48109, United States

[4]Department of Physics, The University of Texas at Dallas, Richardson, Texas 75080, United States

*Corresponding author. Email: dengh@umich.edu

**Two dimensional (2D) van der Waals (vdW) magnetic semiconductors are a new class of quantum materials for studying the emergent physics of excitons and spins in the 2D limit. Twist engineering provides a powerful tool to manipulate the fundamental properties of 2D vdW materials. Here, we show that twist engineering of the anisotropic ferromagnetic monolayer semiconductor, CrSBr, leads to bilayer magnetic semiconductors with continuously tunable magnetic moment, dielectric anisotropy, exciton energy and linear dichroism. We furthermore provide a model for exciton energy in the media with tunable anisotropy. These results advance fundamental studies on 2D vdW materials and open doors to applications to nano-optics, twistronics, and spintronics.**

Two-dimensional (2D) van der Waals (vdW) magnets provide a new platform for studying magnetism in the 2D limit, [1–3] where many-body physics are significantly different from three-dimensional bulk materials and strong coupling of magnetism with other degrees of freedom may emerge. [4] At the same time, unique to 2D vdW materials, twist engineering provides a powerful tool to significantly modify fundamental properties of the material, such as band structures, magnetisms [5–12] and optical responses, giving rise to new states of matters. [13] A particularly interesting material for twist engineering is CrSBr, a 2D vdW magnet that features anisotropic lattice, spin, exciton and optical properties and, remarkably, twist-dependent coupling of these different degrees of freedoms. [14–20] For example, it has been shown that, in a 7-layer CrSBr placed on a twisted grating resonator, anisotropic magnetic polaritons form with unconventional polarization properties; [21] in twisted bilayer (tBL) CrSBr, the spin anisotropy leads to unconventional magnetic transport properties even without interlayer spin coupling. [17,18] However, excitonic and optical properties in twisted CrSBr remain mostly unknown.

Here, we show that, unlike many commonly studied vdW materials, charge tunneling between the two monolayers (MLs) in our tBL CrSBr is negligible. However, we observe strong twist-dependence of total magnetization, dielectric anisotropy, and exciton properties of CrSBr. The results provide insights into twist-dependence of properties in anisotropies of 2D vdW semiconductors and suggest potential optoelectrical and magneto-optical applications.

CrSBr has an orthorhombic structure with spontaneous magnetization following the b-axis in each ML and orienting to opposite directions between adjacent layers (Fig. 1a). [22] Each CrSBr ML is in ferromagnetic order, CrSBr natural bilayer (nBL) and bulk crystal are in antiferromagnetic order. [23] In this work, we make CrSBr tBLs with different twist angles from 0° to 90°, by stacking two CrSBr MLs on 90 nm $SiO_2$/Si substrates, and compare their optical properties (Fig. 1b).

We first investigated how twisting affects the interlayer tunneling in tBLs. It is well established that CrSBr MLs are in-plane ferromagnets (FMs) while natural multilayer CrSBr crystals are interlayer antiferromagnets (AFMs). Inter-layer tunneling in the AFM is forbidden due to anti-aligned spins. An external out-of-plane magnetic field ($B_\perp$) tilts the anti-aligned spins out of plane, which turns on interlayer tunneling and leads to a decrease

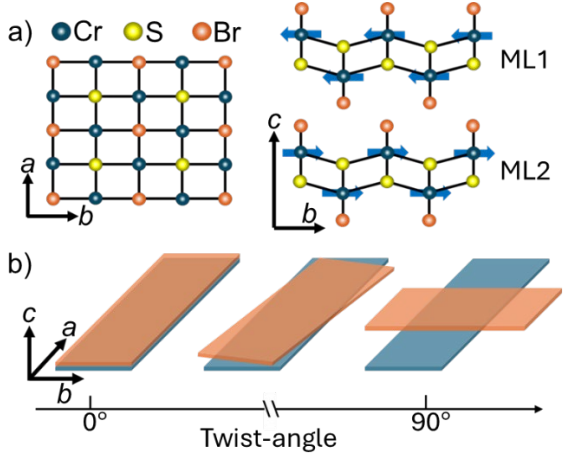

Fig. 1 (a) Scheme of CrSBr lattice with the top view (left panel) and side view (right panel). Blue arrows: spins on Cr atoms. (b) Scheme of CrSBr natural bilayer (nBL) and twisted bilayer (tBL) with twist angle up to 90°.

of the exciton energy in multi-layer CrSBr [14–16] Therefore, $B_\perp$-dependence of the exciton resonance provides a sensitive probe to interlayer tunneling. We measure such dependence via reflection contrast (RC) spectroscopy, the difference between sample (R) and substrate reflection ($R_0$) divided by the substrate reflection: $(R-R_0)/R_0$, for nBL and tBLs under $B_\perp$ from 0 T to 2.4 T, where the exciton resonance appears as a dip in the spectrum (Figs. 2a-d).

As expected, the exciton resonance of the nBL shows a strong B-field dependence. It is centered at $1.331\pm0.002$ eV at 0T and redshifts with increasing $B_\perp$, saturating at $1.322\pm0.002$ eV at $B_\perp \geq 1.8$ T (Figs. 2a&b). This is consistent with the continuous change of the band-gap due to interlayer tunneling, as magnetization of the two layers are aligned by $B_\perp$. [15,16]

In contrast, the RC spectra of the 2°-tBL remains the same under different $B_\perp$(Fig. 2c). Moreover, it has a similar spectral shape as that of the ML, with the same relatively broad exciton resonance centered at $1.308\pm0.002$ eV (Fig. 2d), corresponding to the magnetically confined surface excitons in a CrSBr ML. [24] The depth of the tBL exciton peak is also about twice of that of the ML, suggesting that the 2°-tBL spectra can be considered as a superposition of two MLs' spectra. Similar to the 2°-tBL, tBL of larger twist angles all show similar RC spectral shapes as that of the ML, and their spectra do not shift with $B_\perp$ (see Fig. S2 in Supplementary Materials), as shown in Fig. 2e. These results suggest that, unlike in the nBL, interlayer tunneling between the two MLs in the tBL remain unchanged under $B_\perp$.

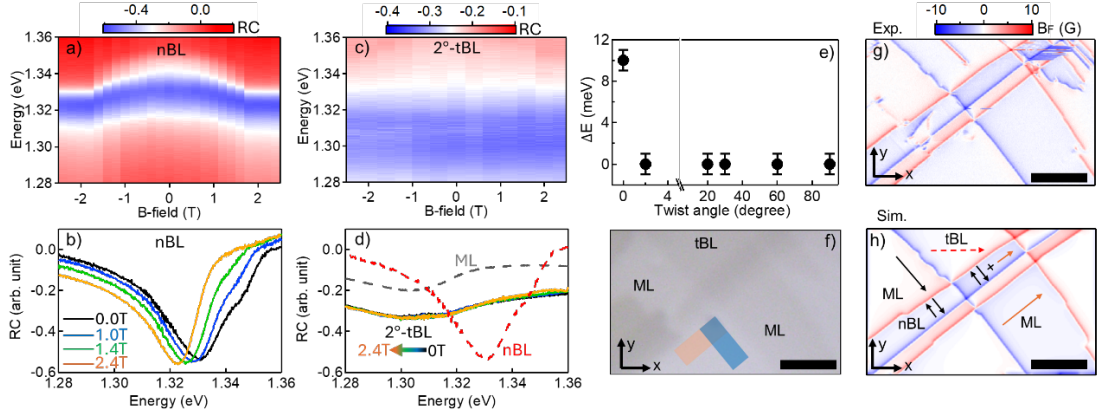

Fig. 2 Magnetic field (B) dependent RC spectra of (a, b) CrSBr nBL and (c, d) 2°-tBL in (a, c) 2D-plot and (b, d) 1D-plot. (b, d) plot spectra at four selected fields (0T, 1T, 1.4T, and 2.4T) from (a, c). Grey (red) dashed line in (d): RC of CrSBr ML (nBL) at 0T. (e) Exciton energy shift (ΔE) between 0 and 2.4T out-of-plane field as a function of twist angle. (f) optical image, (g) magnetic stray field $B_F$ (parallel to NV axis) map, and (h) the corresponding simulation results of CrSBr 90°-tBL. Scale bars in (f-h): 5 μm. Inset of (f): schematic of CrSBr 90°-tBL.

Considering the exciton resonances of CrSBr bilayer is sensitive to the alignment between the spins in the upper- and lower-layer, [15,16] this clear lack of B-field independence of the exciton resonances in tBLs suggests three possibilities: (1) tBL does not form a magnetic order even under an external magnetic field; (2) spins in the two layers are already aligned at zero B-field, so that interlayer tunneling is always allowed and does not change with external magnetic field; (3) the interlayer tunneling is weakened due to the twist between top and bottom layers of tBL, so its effects on the band structure or band-gap becomes negligible.

To check the first possibility, we measure the magnetic order in 90°-tBL (Fig. 2f) using scanning nitrogen vacancy (NV) microscopy (see Methods in Supplementary Materials for experimental details). [11] Fig. 2f and 2g show the optical image and magnetic stray field map (stray field from the magnetic moment of the sample as a function of spatial location), respectively, of tBL and two ML regions of 90°-tBL sample. Significant magnetic stray field are measured, mainly at the edges of the flake, suggesting a uniformly magnetized flake [25,26] and excluding the first possibility. The second possibility is unlikely as natural CrSBr multi-layer is known to form interlayer AFM instead of FM [14] and CrSBr tBL is predicted to have spins in each ML aligned to its own b-axis. [19] Consider the third possibility that magnetizations in both MLs follow their own *b*-axis, in both tBL and ML regions, we obtain the simulated magnetic stray field map as shown in Fig. 2h. The simulation excellently reproduces our experimental map, confirming that the local magnetization in each ML stays in their own magnetic easy states. Based on these results, we exclude the first two possibilities and conclude that the interlayer tunneling is too weak to modify the bandgap in our CrSBr tBL with a twist angle of 2° to 90°. Notably, these results also suggest that the total magnetization of tBL is a superposition of the two MLs, with its magnitude and direction tunable by the twist angle.

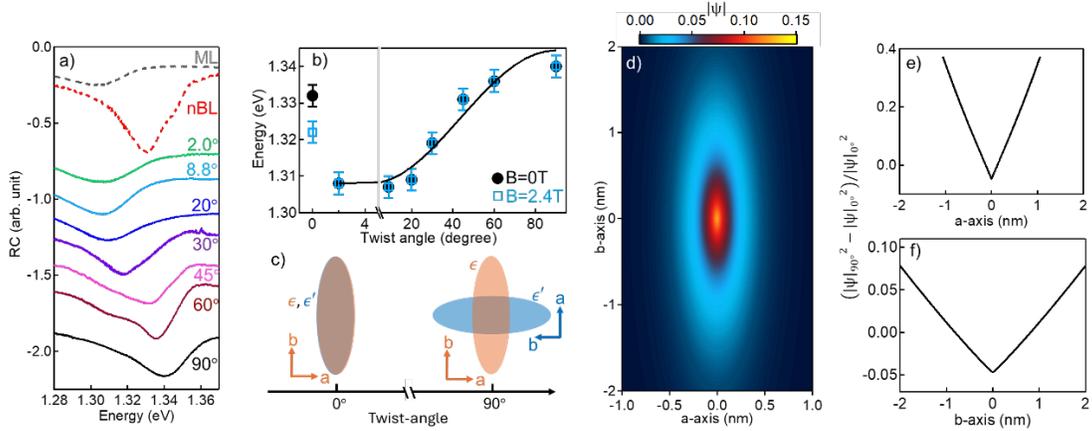

Fig. 3 (a) Twist-angle dependent RC spectra of CrSBr tBLs at 0T. (b) Twist-angle dependent exciton energy of CrSBr tBLs at 0T (black spheres) and 2.4T (blue rectangles). (c) Scheme of dielectric constant of CrSBr tBL with 0° and 90° twist-angle, respectively. (d) 2D plot of normalized in-plane wavefunction of the ground state exciton ($|\psi|$) in CrSBr 90°-tBL. (e-f) Change of the exciton wavefunction between 0° and 90°-tBLs, $\left(|\psi|_{90°}^2 - |\psi|_{0°}^2\right)/|\psi|_{0°}^2$, along (e) a-axis and (f) b-axis, from line-cuts through the center of the exciton wavefunction.

Surprisingly, although there is negligible interlayer tunneling in tBLs, the exciton resonance shows a strong shift with the twist angle. As shown in the RC spectra of CrSBr ML, nBL, and tBLs in Fig. 3a, as the twist angle of the tBL increases from 2° to 90°, the exciton resonance blueshifts from 1.308±0.002 eV to 1.340±0.002 eV (Fig. 3b). Absent of magnetic interlayer coupling, such a shift suggests the exciton resonance results from the twist-dependent dielectric anisotropy in tBLs. Since CrSBr is highly anisotropic, two CrSBr MLs see each other as an anisotropic dielectric environment. Thus, the effective dielectric anisotropy for excitons in tBLs reduces with twist angle, from 0° to 90° (Fig. 3c).

Existing models of 2D excitons all consider media that are isotropic in the 2D plane of exciton motion. Here we develop a model to count for the effect of dielectric anisotropy on the exciton energy in CrSBr tBLs. This model considers the effective dielectric tensor for CrSBr tBLs ($\epsilon_{eff}$) to be a superposition of the two anisotropic dielectric tensors for each CrSBr ML, $\epsilon = \begin{pmatrix} \epsilon_x & & \\ & \epsilon_y & \\ & & \epsilon_z \end{pmatrix}$, with a twist between the two. The twist by an angle θ corresponds to applying a rotation ($R_\theta$). Since the exciton is mainly confined within one of the MLs, we introduced a weight (α) for the effect from the other ML and obtain $\epsilon_{eff} = \frac{\alpha}{1+\alpha} R_\theta \cdot \epsilon \cdot R_\theta^{-1} + \frac{1}{1+\alpha} \epsilon$. Using $\epsilon_{eff}$ for the electric potential, with other material parameters taken from the literature [27,28], we solve the Schrödinger equation of the relative motion of exciton in CrSBr tBL and obtain the exciton binding energy as a function of twist angle θ.

To compare the model with our experimental results, we fit the exciton energy of tBL as a function of twist angle (≥2°) using the reported value of CrSBr ML band gap [27] and with the weight, α, and the dielectric constant along the c-axis of CrSBr ML, $\epsilon_z$, as the fitting parameters (see Methods in Supplementary Materials for details). The model fits very well our experimental data (solid line in Fig. 3b), and the best fit gives $\alpha = 0.04 \pm 0.01$, $\epsilon_z = 1.334 \pm 0.001$, and a binding energy of 802±3 meV. This binding energy is consistent with the previous calculated result. [27] These results reveal that exciton binding energy sensitively depends on the dielectric anisotropy of the medium. Reduced anisotropy leads to a reduced exciton binding energy and a blueshift of the exciton energy.

The effect of dielectric anisotropy on the exciton also manifests in the exciton wavefunction. Fig. 3d is a 2D-plot of the calculated exciton in-plane wavefunction in CrSBr 90°-tBLs from the best fit of the twist-dependent exciton energy (The exciton wavefunction for other twist angles are shown in Fig. S3 in Supplementary Materials). The full width of the half maximum (FWHM) of the normalized exciton wavefunction is 0.12±0.01 nm and 0.44±0.01 nm along a- and b-axis, respectively. This aspect ratio of 3.67 is consistent with a recently calculated ratio of exciton Bohr radii for bulk CrSBr [29] and shows that the quasi-1D nature of intralayer excitons is maintained in the tBL. Fig. 3e (Fig. 3f) shows the normalized difference between the exciton wavefunctions in 0° and 90°-tBLs, $(|\psi|_{90°}^2 - |\psi|_{0°}^2)/|\psi|_{0°}^2$, along the a-axis (b-axis), for line-cuts through the center of the exciton wavefunction. It shows an increase of the size of the exciton as the dielectric anisotropy is reduced. The fractional change of the wavefunction along the a-axis is significantly larger than that along the b-axis. This is because of the much stronger effective confinement of exciton along the a-axis, so the wavefunction along that dimension is more sensitive to the anisotropy change. These results show that the exciton maintains its quasi-1D like, highly anisotropic wavefunction up to a 90° twist angle, but its size increases as the dielectric anisotropy is reduced.

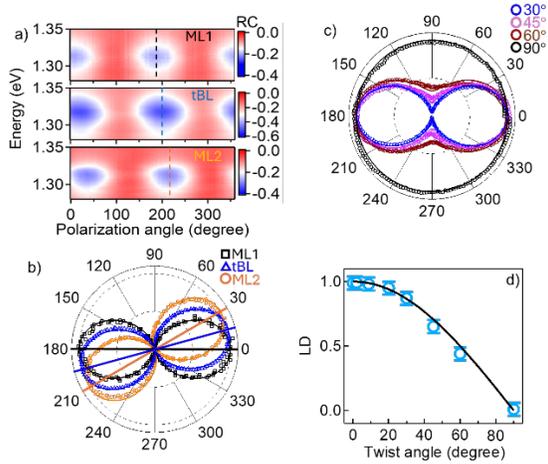

Fig. 4 (a) Polarization-dependent RC spectra of ML1 (upper panel), tBL (middle panel), and ML2 (lower panel) of 30°-CrSBr tBL. (b) Polarization-angle-resolved normalized exciton amplitude of ML1, tBL, and ML2 (symbols) of 30°-tBL. Solid curves: fits. Straight lines: fitted polarization direction. (c) Polarization-angle-resolved normalized exciton amplitude of tBLs with different twist angles. Solid curves: fits. (d) Degree of linear dichroism (LD) of tBL as a function of twist angle (symbols). Solid curve: fit of LD.

Lastly, we study the twist-angle dependence of the optical dichroism of tBLs. Fig. 4a shows the polarization-dependent RC spectra of 30°-tBL (middle panel) and its two component MLs (upper and lower panels). Both MLs have excitons linearly polarized along their respective b-axis, while the tBL has an exciton resonance polarized in between the ML ones (Figs. 4a&b). As we increase the twist angle, the exciton polarization changes from nearly linearly polarized to isotropic (Fig. 4c). We define the degree of linear dichroism as LD = ($I_{max}$ - $I_{min}$)/($I_{max}$ + $I_{min}$), where $I_{max}$ and $I_{min}$ are the maximum and minimum of the polarization dependent exciton peak intensity in RC. The LD varies continuously from near unity at 0°, in nBL and 2°-tBL, to near zero in 90°-tBL, in agreement with a polarization state produced by a sum of those of the two MLs (solid line in Fig. 4d). This result shows that, as the strong Coulomb binding and quasi-1D wavefunction of excitons is maintain in each ML of a tBL, the optical dichroism or exciton polarization become highly tunable by the twist angle.

In summary, we demonstrate continuous tuning of magnetic moment, excitonic energy and dielectric and optical anisotropies in CrSBr tBLs with the twist angle. Interestingly, the tuning results primarily from the anisotropic lattice, spin and exciton properties of CrSBr MLs, while interlayer tunneling is negligible and quasi-1D feature of intralayer excitons is maintained, including strong exciton binding, large exciton oscillator strength, and compact but highly anisotropic exciton wavefunction. We present a new model for excitons in anisotropic 2D media, which describes very well the observed reduction of exciton energy with decreasing dielectric anisotropy in the CrSBr tBLs. These results advance our fundamental understanding of anisotropic 2D vdW materials and open doors to innovative applications to nano-optics, twistronics, and spintronics. For example, optical anisotropy or LD is a fundamental property of crystals and plays a crucial role in optics, photonics, and optoelectronic applications. Twist-controlled CrSBr tBLs form a readily-integrated, ultrathin material with LD continuously tunable between unity and zero.

**Acknowledgements**

Q.L. acknowledges support by DARPA (HR0011-25-3-0317). K.S. and H.D. acknowledge support by the Office of Naval Research (N00014-21-1-2770) and the Gordon and Betty Moore Foundation (GBMF10694). H.D. acknowledges support by the Army Research Office (W911NF-25-1-0055). C.R.D. and S.L. acknowledges support from the U.S. Department of Energy (DOE), Office of Science, Basic Energy Sciences (BES), under award No. DE-SC0024870. The work at the University of Texas at Dallas are supported by US Air Force Office of Scientific Research Grant No. FA9550-19-1-0037, National Science Foundation- DMREF-2324033, and Office of Naval Research grant no. N00014-23-1-2020 and N00014-22-1-2755. L. Z. acknowledges support from the U.S. Department of Energy (DOE), Office of Science, Basic Energy Science (BES), under award No. DE-SC0024145.## References

[1] M. Gibertini, M. Koperski, A. F. Morpurgo, and K. S. Novoselov, Magnetic 2D materials and heterostructures, Nat. Nanotechnol. **14**, 408 (2019).

[2] K. S. Burch, D. Mandrus, and J.-G. Park, Magnetism in two-dimensional van der Waals materials, Nature **563**, 47 (2018).

[3] K. F. Mak, J. Shan, and D. C. Ralph, Probing and controlling magnetic states in 2D layered magnetic materials, Nat Rev Phys **1**, 646 (2019).

[4] Y. Ahn, X. Guo, S. Son, Z. Sun, and L. Zhao, Progress and prospects in two-dimensional magnetism of van der Waals materials, Progress in Quantum Electronics **93**, 100498 (2024).

[5] H. Xie et al., Twist engineering of the two-dimensional magnetism in double bilayer chromium triiodide homostructures, Nat. Phys. **18**, 30 (2022).

[6] T. Song et al., Direct visualization of magnetic domains and moiré magnetism in twisted 2D magnets, Science **374**, 1140 (2021).

[7] Y. Xu et al., Coexisting ferromagnetic–antiferromagnetic state in twisted bilayer CrI3, Nat. Nanotechnol. **17**, 143 (2022).

[8] H. Xie et al., Evidence of non-collinear spin texture in magnetic moiré superlattices, Nat. Phys. **19**, 1150 (2023).

[9] M. Huang et al., Revealing intrinsic domains and fluctuations of moiré magnetism by a wide-field quantum microscope, Nat. Commun. **14**, 5259 (2023).

[10] G. Cheng et al., Electrically tunable moiré magnetism in twisted double bilayers of chromium triiodide, Nat. Electron. **6**, 434 (2023).

[11] S. Li et al., Observation of stacking engineered magnetic phase transitions within moiré supercells of twisted van der Waals magnets, Nat. Commun. **15**, 5712 (2024).

[12] B. Yang et al., Macroscopic tunneling probe of Moiré spin textures in twisted CrI3, Nat. Commun. **15**, 4982 (2024).

[13] D. M. Kennes, M. Claassen, L. Xian, A. Georges, A. J. Millis, J. Hone, C. R. Dean, D. N. Basov, A. N. Pasupathy, and A. Rubio, Moiré heterostructures as a condensed-matter quantum simulator, Nat. Phys. **17**, 155 (2021).

[14] N. P. Wilson et al., Interlayer electronic coupling on demand in a 2D magnetic semiconductor, Nat. Mater. **20**, 1657 (2021).

[15] Y. J. Bae et al., Exciton-coupled coherent magnons in a 2D semiconductor, Nature **609**, 282 (2022).

[16] G. M. Diederich et al., Tunable interaction between excitons and hybridized magnons in a layered semiconductor, Nat. Nanotechnol. **18**, 23 (2023).

[17] C. Boix-Constant, S. Jenkins, R. Rama-Eiroa, E. J. G. Santos, S. Mañas-Valero, and E. Coronado, Multistep magnetization switching in orthogonally twisted ferromagnetic monolayers, Nat. Mater. **23**, 212 (2024).

[18] Y. Chen, K. Samanta, N. A. Shahed, H. Zhang, C. Fang, A. Ernst, E. Y. Tsymbal, and S. S. P. Parkin, Twist-assisted all-antiferromagnetic tunnel junction in the atomic limit, Nature **632**, 1045 (2024).
9